# Spatio-temporal modeling of yellow taxi demands in New York City using Generalized STAR models

Abolfazl Safikhani [1], Camille Kamga, Sandeep Mudigonda, Sabiheh Sadat Faghih, Bahman Moghimi

as5012@columbia.edu, ckamga@utrc2.org, mudigonda@utrc2.org, sfaghih00@citymail.cuny.edu, smoghim000@citymail.cuny.edu

**Abstract**

*A highly dynamic urban space in a metropolis such as New York City, the spatio-temporal variation in demand for transportation, particularly taxis, is impacted by various factors such as commuting, weather, road work and closures, disruption in transit services, etc. To understand the user demand for taxis through space and time, a generalized spatio-temporal autoregressive (STAR) model is proposed in this study. In order to deal with the high dimensionality of the model, LASSO-type penalized methods are proposed to tackle the parameter estimation. The forecasting performance of the proposed models is measured using the out-of-sample mean squared prediction error (MSPE), and it is found that the proposed models outperform other alternative models such as vector autoregressive (VAR) models. The proposed modeling framework has an easily interpretable parameter structure and practical to be applied by taxi operators. Efficiency of the proposed model also helps in model estimation in real-time applications.*

[1]Assistant Professor, Department of Statistics, Columbia University in the City of New York (corresponding author)

# 1. Introduction

Taxi services have been an important part of urban transportation. Traditionally, ride hailing is performed by customers on the curbside of streets. The street hail taxis could be inefficient in addressing spatiotemporal variations in demand. Given the highly dynamic urban space in a metropolis such as New York City, the spatiotemporal variation in demand for taxis is impacted by various factors such as commuting, weather, special events, parades, road work and closures, disruption in transit services, etc. However, with the rise of transportation network companies (TNCs) ride hailing is becoming more on-demand. TNCs use economic means such as surge pricing to address some inefficiency. Instead, accurate prediction of the demand for taxis could lead to much better efficiency and more nuanced economic measures. Additionally, accurate short-term demand prediction for taxis enables the TNCs to dynamically reroute, schedule and optimize operations.

The demand for ride hailing taxis in New York City (NYC) is highly variable with a maximum of about 600,000 to a minimum of about 150,000 trips per day provided by 21,263 street hail taxis in 2015 (TCL Factbook, 2016) (as seen in Figure 1). This demand also has a high spatial variability with about 383,000 pickups in Manhattan and only 3,150 pickups in the Bronx on an average day. GPS enabled spatio-temporal historical demand for taxis in the year of 2015 to be disaggregated to several sub-regions within the city.

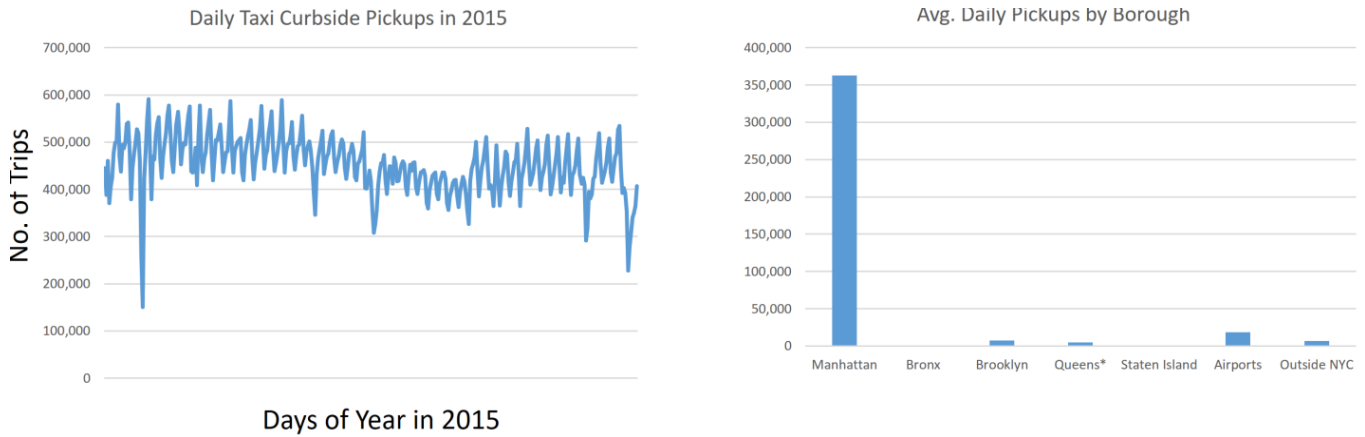

*Figure 1 Temporal and Spatial variation in taxi demand*

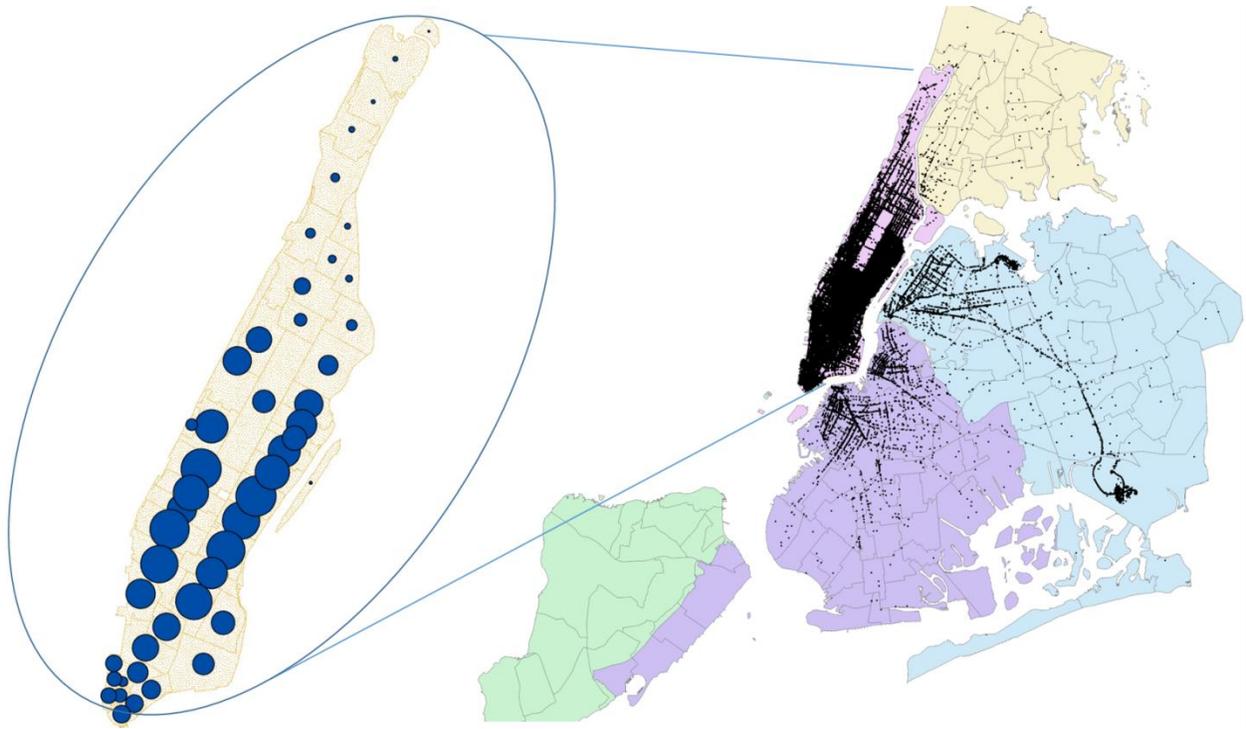

*Figure 2 Spatial variation of taxi demand aggregated by zipcode in Manhattan*

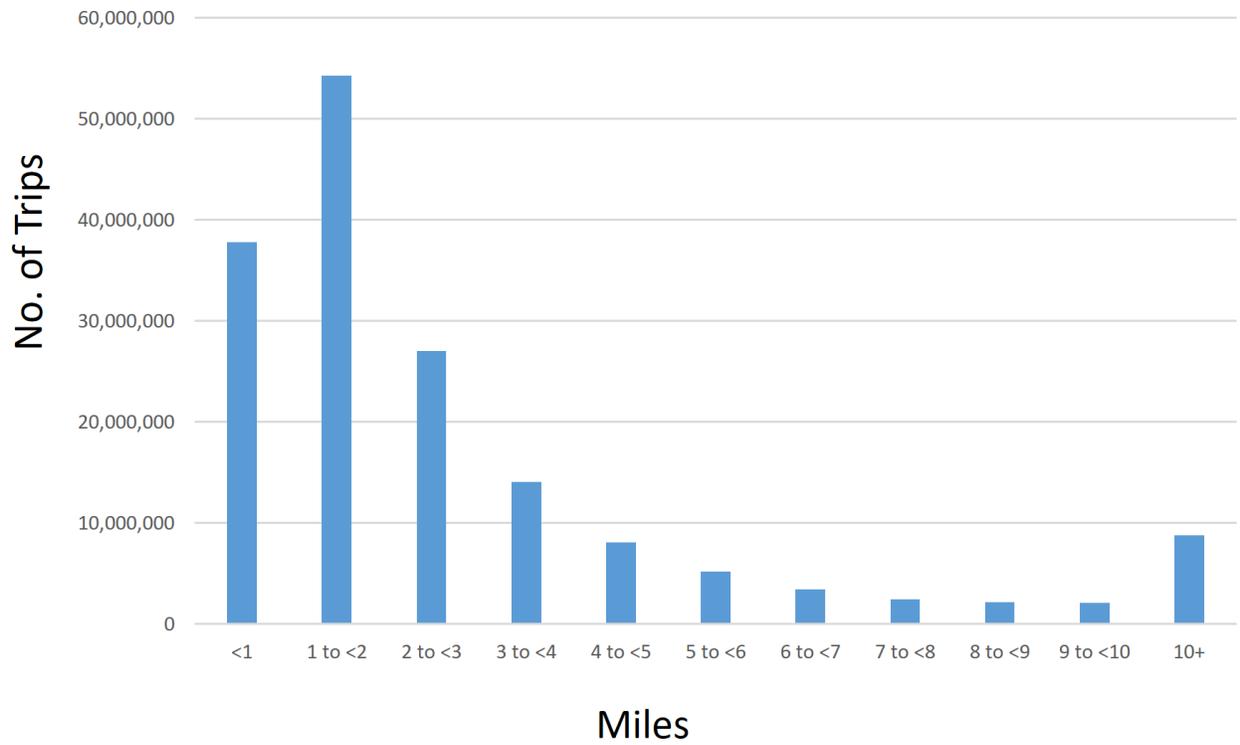

*Figure 3 Distribution of distance travelled by taxis in 2015*

These taxis travelled approximately 460 million miles in 2015 (TLC Factbook, 2016) the distribution of which can be seen in Figure 3. Due to the myriad factors impacting demand, which may or may not be known in advance, there is scope for taxis driving around seeking rides – some of which could be in the number of trips under one mile in Figure 3.

In this study, demand for taxi is modeled as a dynamic spatio-temporal process. GPS-enabled spatio-temporal historical demand for taxis in the year of 2015 (provided by the Taxi and Limousine Commission of New York City) is used and aggregated to several sub-regions within the city.

There were some studies to prepositioning taxis for reduced wait time (Chan et al. (2010), Yuan et al. (2011) using spatiotemporal clustering. Time series models such as ARIMA has also been tested for predicting taxi demand prediction (Moreira-Matias et al. (2013), Sayarshad and Chow (2016) and Qian et al. (2017). Artificial neural networks were also applied to combat nonlinearities in tax demand (Qian et al., 2017). Furthermore, spatio-temporal variations were attempted to be captured using conditional random fields. (Qian et al., 2017).

In order to understand the demand's behavior through space and time, we use a spatio-temporal ARMA (STARMA) model. STARMA model is a well-established spatio-temporal process introduced by Pfeifer & Deutrch (1980 & 1981), and it has been applied in many different disciplines such as social science (Pfeifer and Deutsch (1980) and Sartoris (2005)), transportation (Kamriankis and Prastacos, 2003; Cheng et al., 2011; Duan et al., 2016), climatology (Kyriakidis and Journel, 1999), economics (Giacomini and Granger, 2004), health sciences (Baklanov et al., 2007), etc. Modeling the demand through time in all the sub-regions simultaneously is a high-dimensional problem since the number of parameters in the model is proportional to the squared of the number of sub-regions. STARMA reduces the number of parameters dramatically by governing a neighborhood structure between the regions. This structure is also useful in capturing the spatial dependence of the demand between the regions and further makes the results more interpretable. To evaluate the performance of the proposed method, the forecasting performance of such model is measured using the out-of-sample mean squared prediction error (MSPE), and the results have shown that the proposed model has outperformed some alternative algorithms such as ARMA and VAR models.

Given that there are about 12 million taxi trips a month that amounts to 2 GB of data, a demand forecasting model with accurate spatial and temporal predictability is very useful. Particularly, the proposed model has the ability to forecast the taxi demand few steps ahead in the future at various locations in NYC, and this enables the agencies for the real-time provision of demand-sensitive taxi dispatching for various locations and specific times of the day over the year. This is particularly useful for the operating agency so that empty ride-seeking taxi trips and thus the fuel burned can be lowered. Such demand-sensitive dispatch also has an environmental benefit by reducing the emissions associated to empty ride-seeking taxi trips. Additionally, from a policy standpoint, the spatio-temporal structure inferred from the demand data provides a basis for regulating agencies to explore cordon pricing initiatives.

This paper is organized as follows; the second section discusses the literature about time series modeling in transportation and short-term taxi demand prediction. The third section describes, in detail, the spatiotemporal modeling and formulation of taxi demand using STARMA approach. The fourth section presents findings for various types of STARMA models and prediction errors and compares with other time series models. Finally, we present the conclusions and future research directions.

## 2. Literature Review

With the rise of intelligent system and availability of data, taxi pick-up demand prediction has recently come into attention of many scholars. It is explicit that taxi demand of each zone changes from one time interval to another in real time. Hence, to capture this time variation in taxi demand, figuring out the correlation among the taxi data, and having real time prediction, time series model can be a strong statistical tool. For a univariate data, a well-known family of time series called ARIMA (Autoregressive Integrated Moving Average) can be beneficial, and it has been applied to many transportation related problems (Moghimi et al., 2017). However, in a dense urban transportation network which there are many areas or zip-codes that each has its own demand dynamic and possibly correlated to one another, the taxi demand variation of such zip-code is not just related to its own values, whereas it is affected by demand of the neighboring zip-codes. Since there are too many parameters to estimate due to having multiple zip-codes, VAR (Vector Autoregressive) models, as the most common multivariate time series models, will not be able to perform well and will make poor forecasting of taxi demand. To mitigate this problem, a spatial-temporal time series model as a family of multivariate time series model, is being applied in this study.

Some of the primary researches about taxi demand were to find factors influencing taxi demand. Schaller (1999) developed a citywide empirical time series regression model on NYC taxi to understand the relationship between taxicab revenue per mile and economic activity in the city, taxi supply, taxi fare, and bus fare. Afterward, Schaller (2005) tried to figure out the relationships between taxi demand and factors including city size, availability and cost of privately owned autos, use of complements to taxicabs, cost of taxi usage, taxi service quality, presence of competing modes, senior and disabled population.

Subsequently and with the emergence of GPS technology, extensive researches about spatial information have been applied in transportation-related problems. GPS based system is also utilized on taxis of New York City to track them and to analyze taxi ridership with such data source. Yang and Gonzales (2017) processed the GPS taxi data of New York City and used negative binomial method to capture the variation of taxi pick-up demand. Six explanatory variables were used in their study including population, education, median age, median income per capita, employment by industry sector, and transit accessibility. Correa et al. (2017) performed empirical analysis to explore the spatial dependence between Uber and taxi pick-up data. Results from Moran's I tests confirmed the significantly spatial correlation of both taxi and Uber demand.

Moreira-Matias et al. (2013) proposed a methodology to predict short-term taxi demand at 30 min time intervals. Their methodology is an ensemble of three predictive models including Time Varying Poisson Model, Weighted Time Varying Poisson Model, and Auto Regressive Integrated Moving Average Model. They found that their proposed model outperformed to all three models if run individually. In the recent study done by Qian et al. (2017), a Gaussian Conditional Random Field (GCRF) model is presented to predict a short-term taxi demand. The proposed model together with 2 other algorithms (ARIMA and ANN) were run in 4 different scenarios to evaluate its performance. The results reported that the proposed model outperformed the two other algorithms with Mean absolute percentage error (MAPE) close to 0.1. In this paper, the case study was the same as used in Qian et al. (2017) study.

It is well-proven that spatial information increases the accuracy of prediction specifically in congestion traffic and for longer horizon. The idea of capturing spatial information in the times series studies of transportation related problems was firstly introduced in the study by Okutani and Stephanedes (1984) to predict traffic flow prediction. The spatial concept later was deployed in the study by Kamriankis and Prastacos (2003) to forecast the relative velocity on major roads in Athens, Greece. They called the method space-time autoregressive integrated moving average (STARIMA). The model is quite different from traditional ARIMA model by including the spatial information of neighboring links for traffic forecasting. They compared the forecasting performance in four models including historical average, ARIMA, VARMA, and STARIMA. The results demonstrated that there is no significant difference between the last three models although the last three models performed better than the historical average one. Using spatial-temporal modeling is also used in other areas of transportation. For example, the traffic condition of downstream section of a road is highly correlated to the traffic condition coming from upstream. Stathopoulos & Karlaftis (2003) considered the spatial information of four consecutive loop detectors from the upstream of the study section to predict the traffic flow in the downstream of an urban corridor. The same idea was used in the study done by (Cheng et al., 2011; Duan et al., 2016) to predict traffic speed of downstream link.

In STARIMA modeling, the spatial weighting matrix is one of the most important parts which is related the spatial dependency between multiple time series. Thus, how to make the spatial weighting matrix varies by the nature of each problem, and it needs some engineering judgment. Mostly, two approaches have been used to select the neighboring dependence: (a) correlation-coefficient assessment and (b) distance adjustment. The values in STARIMA's weighting matrix can vary by time and location. In a developed method called General STARIMA, the spatial parameters are designed to vary per location instead of having fixed values over all locations (Min et al., 2010). In Dynamic STARIMA model, which was presented by Min et al. (2009), a practical approach was used to forecast short-term traffic flow in urban road network in Beijing, China. In the developed Dynamic STARIMA model, instead of having a static weighting matrix, a dynamic weighting matrix is used that its values change from time to time depending on time-varying lag of the upstream time points. In their study, the matrix's values represent the proportion of volume form upstream intersection affecting the downstream link's flow. For instance, to forecast the flow of link at downstream intersection, the proportion of flow that turn right, left, and through from upstream affecting the downstream intersection are used; and the turning-values are not fixed anymore whereas they are estimated from the previous time lag. Another approach that associated with weighting matrix is to just consider link/zone that is adjacent to the target link/zone. It can be elaborated by ring of dependency as labeled by "order". For instance, first-order adjacent matrix represents the dependency between the study link/zone to its immediate adjacent link/zone. Second-order adjacent matrix shows the zone is indirectly close to the study zone but having direct dependency to the link/zone defined as first-order. It can expand to third-order adjacent matrix, and so forth. First and second order adjacency-weighting matrix was used in the study done by Kamarianakis et al. (2004). On the other hand, it is more practical to use the distance between the two links/zones, which the value of dependency reduces by increasing the distance.

## 3. Methodology

In this section, the proposed spatiotemporal model will be introduced and the implementation of the model will be briefly discussed. Suppose $k$ different time series data are observed over duration of size $T$. If one chooses vector auto regression (VAR) models with max time lag being $p$ to fit the data, it means in total $k^2 * p$ parameters need to estimated using the $k * T$ total observed data points. Now, if $k$ is relatively large as compared to $T$, then the number of parameters in the model will be more than the observed data. This is called a high-dimensional problem. The typical least square methods cannot be used as the design matrix will not be invertible. The data set we are exploring in this paper, shares similar features to high-dimensional time series. More specially, the yellow taxi demand in NYC is considered for the day October 6th, 2015. The reason this date is chosen is that it is a typical day without any holidays or any special events nearby. Then, the demand is aggregated spatially over the zip-codes, and temporally every 15 minutes. Therefore, it is a multivariate time series with more than 100 components. However, only 39 of the zip-codes have enough non-zero counts to keep them in the model. Thus, finally the data consists of $k = 39$ locations, and $T = 96$ time points. Figure 4 shows the sample ACF of the first 5 components of the data which implied existence of the strong temporal dependence. Hence, a multivariate time series model is chosen to analyze this data.

*Figure 4:* Sample ACF of the first 5 components

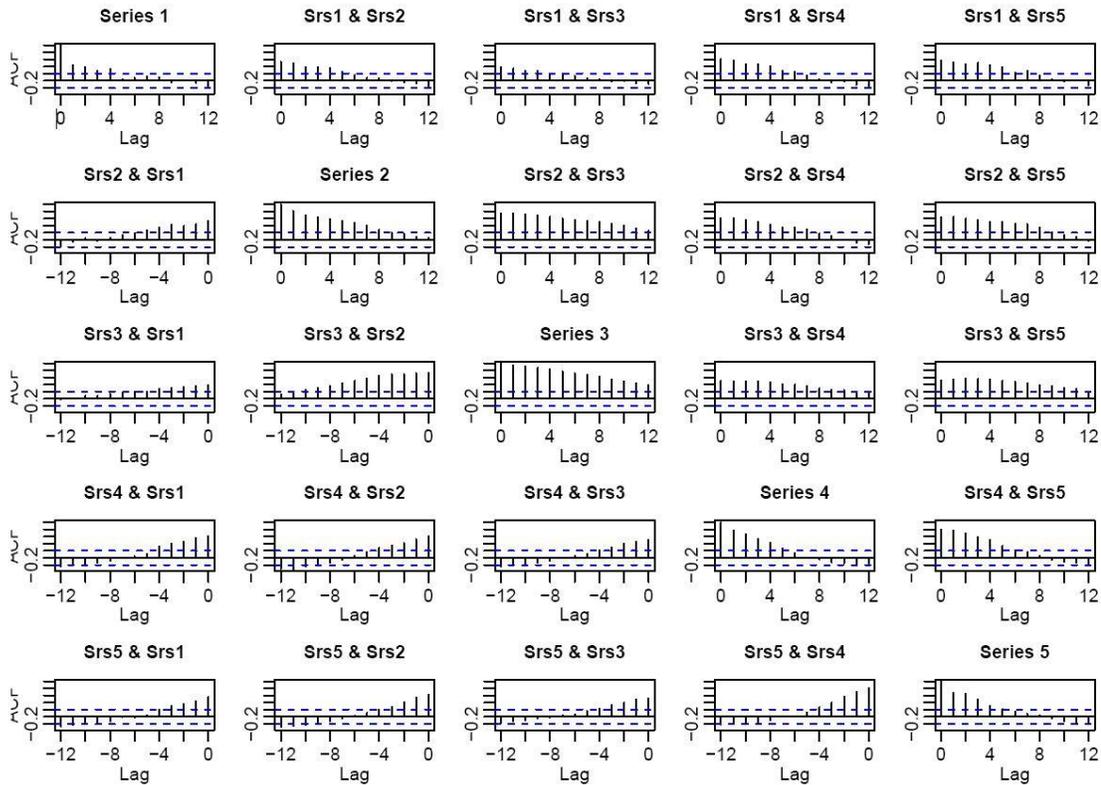

Due to the high-dimensionality of the data, simple VAR models will not be appropriate for the data. Instead, a generalized version of STARMA model, which takes into account the topology of the locations at which the data is observed, is developed in this section with the aim of prediction performance efficiency. STARMA models, introduced by Pfeifer & Deutrch (1980 & 1981), is in general a spatio-temporal model. This model reduces the number of parameters in a typical VAR model by introducing neighborhood structures. Here we only focus on the autoregressive (AR) part of this model since it is more interpretable. A multivariate time series $Y(t) = (Y_1(t), ..., Y_k(t))$, $t = 1, 2, ..., T$ is called to be generalized STAR of order $p$ (See references (Giacinto, 1994; Terzi, 1995) for introduction, and (Giacinto, 2006) for its application to regional unemployment analysis) if for each $t = 1, 2, ..., T$ and $i = 1, 2, ..., k$,

$$Y_i(t) = \sum_{j=1}^{P} \sum_{l=0}^{\eta_j - 1} \phi_i^{(j,l)} W_i^{(l)} Y(t-j) + \varepsilon_i(t), \quad (1)$$

where $\varepsilon_i(t) = (\varepsilon_1(t), ..., \varepsilon_k(t))$ is a k-variate normal variable with mean zero and

$$\mathbb{E}\left(\varepsilon(t)\varepsilon(t+s)'\right) = \begin{cases} \sigma^2 I_k, & s = 0 \\ 0, & otherwise \end{cases}$$

Also, $W^{(l)}$'s are $k * k$ weighting matrices which govern the $l$-th neighborhood location with $W_i^{(0)} = I_k$. Denote the $i$-th row of $W^{(l)}$ by $W_i^{(l)}$. Possible choice for $W^{(l)}$ is to put $W^{(l)}(i,j) = 1$ if $i$-th and $j$-th locations are $l$-th level neighborhood, and $W^{(l)}(i,j) = 0$ otherwise. These matrices are then normalized in such a way that the sum of each row would be 1. Finally, for each $i = 1, 2, ..., k$, and $j = 1, 2, ..., p$, $\phi_i^{(j, 0:\eta_j - 1)} = \left(\phi_i^{(j,0)}, \phi_i^{(j,1)}, ..., \phi_i^{(j, \eta_j - 1)}\right)$ is a vector of coefficients of size $\eta_j$ relating the current observation at location $i$, $Y_i(t)$, to the all weighted observations in $\eta_j$ different neighborhoods $j$ time lags in the past. Without loss of generality, it is assumed that $\eta_1 = \cdots = \eta_p = \eta$ (If they are different, one can choose $\eta = \max(\eta_1, ..., \eta_p)$ and set some of the $\phi_i^{(j,l)}$ coefficients to zero). Further, denote $\Phi_i = \left(\phi_i^{(1, 0:\eta - 1)}, ..., \phi_i^{(p, 0:\eta - 1)}\right)$. It would be more convenient to write equation (1) in a compact matrix form. For that, let $Y_i = Y_i(1), ..., Y_i(T))$, $\varepsilon_i = (\varepsilon_i(1), ..., \varepsilon_i(T))$, and define $Z_i$ to be the $T * \eta p$ with $Z_i(t, (j-1)\eta + l) = W_i^{(l)} Y(t-j)$ for $t = 1, 2, ..., T$, $j = 1, 2, ..., p$, and $l = 0, 2, ..., \eta - 1$. Now, one can write the data equation for $i$-th time series component as follows:

$$Y_i = Z_i \Phi_i + \varepsilon_i \quad (2)$$

This model reduces the number of parameters from $k^2 * p$ in the VAR model to $k * \eta * p$, assuming $\eta \ll k$. Least squares estimation can be implemented for parameter estimation, i.e. for $i = 1, 2, ..., k$,

$$\widehat{\Phi}_i = argmin_{\Phi_i} \frac{1}{2} \|Y_i - Z_i \Phi_i\|_2^2, \quad (3)$$

with $\|.\|_2$ being the Euclidean norm. However, for the cases when $T$ is small compared to $k$, it might be beneficial to still reduce the number of parameters in the model with the goal of improving forecast performance. For that, a penalty function $\Omega(\Phi)$ will be added to equation (3) with the purpose of setting some of the small parameters to zero to increase forecast efficiency. More specifically,

$$\widehat{\Phi}_i = argmin_{\Phi_i} \frac{1}{2} \|Y_i - Z_i \Phi_i\|_2^2 + \lambda \, \Omega(\Phi_i), \quad (4)$$

where $\lambda$ is the tuning parameter to be selected by cross validation techniques. Several penalty functions will be defined, and their performance will be evaluated on the yellow taxi demand data. More specifically, the following penalty functions are considered:

- **LASSO:** Simple element-wise $L_1$ penalty on all the components of $\Phi_i$, i.e. for $i = 1, 2, \ldots, k$,

$$\Omega(\Phi_i) = \sum_{j=1}^{p} \sum_{l=0}^{\eta-1} \left|\phi_i^{(j,l)}\right| \qquad (5)$$

- **HGLASSO** (Hierarchical Group LASSO): This method is similar to the HVAR method introduced in (Nicholson et al., 2014; Nicholson et al., 2017) for sparse VAR models. The coefficients for each time lag are being grouped together, and they are penalized more if the time lags are higher through a time-lag hierarchical group structure. More specifically, denoting $\Phi_i^{(j:p)} = \left(\phi_i^{(j,0:\eta-1)}, \ldots, \phi_i^{(p,0:\eta-1)}\right)$ for $j = 1, 2, \ldots, p$,

$$\Omega(\Phi_i) = \sum_{j=1}^{p} \left\|\Phi_i^{(j:p)}\right\|_2 \qquad (6)$$

- **DHGLASSO** (Double Hierarchical Group LASSO): We propose this penalty function that is similar to HGLASSO, but with an additional neighborhood-lag hierarchical group structure penalty term. Denoting

$$\Phi_i^{(j:p,l:\eta-1)} = \left(\phi_i^{(j,l:\eta-1)}, \phi_i^{(j+1,0:\eta-1)}, \ldots, \phi_i^{(p,0:\eta-1)}\right), \; j = 1, 2, \ldots, p, \; l = 0, 2, \ldots, \eta - 1, \qquad (7)$$

one can write the penalty function as follows:

$$\Omega(\Phi_i) = \sum_{j=1}^{p} \sum_{l=0}^{\eta-1} \left\|\Phi_i^{(j:p,l:\eta-1)}\right\|_2 \qquad (8)$$

### 3.1. Implementation

Solving optimization problems of type (4) has been studied well under the penalty terms introduced previously (See (Tibshirani, 1996) and references therein). Due to the hierarchy structure of the group penalties in HGLASSO and DHGLASSO, here we apply the proximal gradient method introduced in (Jenatton et al., 2011). Further, the convergence rate of the proximal gradient method has been improved in (Beck and Teboulle, 2009) by introducing the Fast Iterative Soft-Thresholding Algorithm (FISTA). In FISTA, a sequence of matrix coefficients $\widehat{\Phi}_i[r]$, $r = 1, 2, \ldots$ are introduced iteratively through

$$\hat{\phi} = \widehat{\Phi}_i[r-1] + \frac{r-2}{r+1}\left(\widehat{\Phi}_i[r-1] - \widehat{\Phi}_i[r-2]\right)$$
$$\widehat{\Phi}_i[r] = Prox_{s\lambda\Omega}\left(\hat{\phi} - s\nabla f_i(\hat{\phi})\right), \qquad (9)$$

with $f_i(\Phi_i) = \frac{1}{2}\|Y_i - Z_i \Phi_i\|_2^2$, $\nabla f_i(\Phi_i) = -Z_i'(Y_i - Z_i \Phi_i)$ the vector of derivatives of $f_i(\Phi_i)$, $s$ being the step-size (here we choose s to be $1/\sigma_1(Z_i)^2$ where $\sigma_1(Z_i)$ is the largest singular value of $Z_i$), and

$$Prox_{s\lambda\Omega}(u) = argmin_v \left(\frac{1}{2}\|u - v\|^2 + s\lambda\Omega(v)\right). \qquad (10)$$

The proximal function may not have a closed form in general, and in that case, it needs to be approximated numerically itself. However, in the case of hierarchical group penalty, this function, in fact, has a simple closed form (See for example algorithm 2 in (Nicholson et al., 2014)). This makes the whole optimization efficient. The tuning parameter $\lambda$ is selected based on a rolling scheme cross-validation procedure used also in (Song and Bickel, 2011; Nicholson et al., 2014; Nicholson et al.; 2017). For this, the time points are divided into three parts (usually equally distanced) $0 < T_1 < T_2 < T$. The estimation procedure for fixed values of $\lambda$ will be applied for the first part, i.e. $t = 1, 2, .., T_1$. Then, the mean squared prediction error (MSPE) for predicting one step ahead is calculated over all $k$ time series components on the time interval $[T_1 + 1, T_2]$:

$$MSPE = \frac{1}{k(T_2-T_1)} \sum_{i=1}^{k} \sum_{t=T_1+1}^{T_2} (Y_i(t) - P_{T_1} Y_i(t))^2 \, , \qquad (11)$$

$$MRPE = \frac{1}{k(T_2-T_1)} \sum_{i=1}^{k} \sum_{t=T_1+1}^{T_2} \left| \frac{Y_i(t) - P_{T_1} Y_i(t)}{Y_i(t)} \right|, \qquad (12)$$

where $P_{T_1} Y_i(t)$ is the best linear predictor of $Y_i(t)$ based on the first $T_1$ observations. Mean of error prediction error (MRPE) is also shown in equation (12). Now, the tuning parameter $\lambda$ which is minimizing this MSPE will be selected, and the model performance then can be quantified by the MSPE on the last part of the data, which is on the time interval $[T_2 + 1, T]$.

## 4. Results

In this section, the proposed methods are applied over the yellow taxi demand data in different days, and their prediction performance is calculated under different scenarios. Based on the sample ACFs of the data, $p$ is chosen to be 1. Also, the calculation on the AIC/BIC supports this selection. Before applying different methods to this data, it needs to be scaled properly. For that purpose, for each time series corresponding to a zip-code, the sample mean is subtracted and then divided by the sample standard deviation so that time series' have same scales. Also, the weighting matrices $W's$ are chosen for five different neighborhood levels based on authors' judgment, more specifically, by counting the number of boundaries between the target zip-code and its neighbors. For example, a zip-code adjacent to the target zip-code is considered as the first order neighborhood; zip-codes adjacent to the first neighborhood order will be a second order neighborhood for the target zip-code and so on. The levels of neighborhood in this study are extended through an eyeballing procedure up to five levels. October 6th and 7th are chosen for this research because of being a typical weekday, being away from a weekend day or a day with special event. Two approaches have been considered to evaluate the performance of the developed model. First is to consider time points of only October 6$^{th}$, and in the second approach two days of October 6$^{th}$ and 7$^{th}$ are merged to have a longer range of time points.

### 4.1. Case Study using Data for October 6$^{th}$ only

Considering data on October 6$^{th}$, only T = 96 time points are available. Rolling scheme method is applied to divide the data in the time series. It means T is divided into 3 parts, setting T$_1$ to be $\lfloor T/3 \rfloor$, and T$_2$ to be $\lfloor 2T/3 \rfloor$. Different orders of neighborhood ($\eta$) are chosen, and the MSPE, mean squared relative prediction error (MRPE), AIC and BIC (See (Lutkepohl, 2007) for the definition and the formula) are reported for each case. Tables 1, 2, 3, and 4 show the results for $\eta = 1, 2, 3, 4$, respectively. Obviously, the VAR model does not perform well due to the huge number of parameters involved as compared to STAR-based models. Based on the MSPE, STAR and LASSO models for $\eta = 2$ are outperforming the rest. This means including first neighborhood structure improves the forecasting performance of the STAR model. Meanwhile, the spatio-temporal structure developed using the topology and zip-code-based disaggregation of Manhattan, the proposed model with first order neighborhood performs the best in this case study. Also, it is worth mentioning that the DHGLASSO penalty function provides consistent model performance overall since its MSPE/MRPE are not increasing dramatically by increasing $\eta$. In other word, DHGLASSO penalty structure corrects better for the increase in the parameter space dimension. Also, by increasing the number of neighborhood levels $\eta$, which means an increase on the number of parameters in $\Phi's$, DHGLASSO method is able to reduce the MSPE as compared to the STAR around 3% when $\eta = 3$, and around 17% when $\eta = 4$. If the MRPE is selected as the forecasting performance measurement, then DHGLASSO when $\eta = 4$ is comparable to the other leading models.

*Table 1: MSPE for October 6$^{th}$ data with $\eta = 1$*

| Model | MSPE | MRPE | AIC | BIC |
|---|---|---|---|---|
| VAR | 1.7153 | 4.8259 | 216.4933 | 257.1222 |
| STAR | 0.2815 | 2.4463 | 176.9854 | 178.0271 |
| LASSO | 0.2977 | 1.8467 | 173.8735 | 174.9153 |
| HGLASSO | 0.2977 | 1.8467 | 173.8735 | 174.9153 |
| DHGLASSO | 0.2977 | 1.8467 | 173.8735 | 174.9153 |

*Table 2: MSPE for October 6$^{th}$ data with $\eta = 2$*

| Model | MSPE | MRPE | AIC | BIC |
|---|---|---|---|---|
| STAR | 0.2707 | 1.9913 | 177.3313 | 179.4148 |
| LASSO | 0.2728 | 1.9616 | 177.0052 | 179.0353 |
| HGLASSO | 0.2909 | 1.8942 | 176.0614 | 178.1449 |
| DHGLASSO | 0.2907 | 1.9543 | 178.6925 | 180.6425 |

Table 3: MSPE for October 6$^{th}$ data with $\eta = 3$

| Model | MSPE | MRPE | AIC | BIC |
|---|---|---|---|---|
| STAR | 0.2932 | 2.1346 | 178.7531 | 181.8784 |
| LASSO | 0.3254 | 2.1413 | 177.3218 | 179.1115 |
| HGLASSO | 0.301 | 2.114 | 175.1811 | 178.3064 |
| DHGLASSO | 0.2821 | 1.9472 | 176.3991 | 179.3107 |

Table 4: MSPE for October 6$^{th}$ data with $\eta = 4$

| Model | MSPE | MRPE | AIC | BIC |
|---|---|---|---|---|
| STAR | 0.3582 | 2.4261 | 182.2474 | 186.3877 |
| LASSO | 0.3506 | 2.2577 | 177.8353 | 180.3196 |
| HGLASSO | 0.3412 | 2.2928 | 177.0926 | 181.2329 |
| DHGLASSO | 0.2968 | 1.882 | 176.5145 | 180.0939 |

## 4.2. Case Study using Data for October 6$^{th}$ and 7$^{th}$ Combined

The same set of models and methods applied in the previous case study are applied using the taxi demand for two days, October 6$^{th}$ and 7$^{th}$. This makes the total number of time points to be 192 instead of 96 as in previous case study. Increasing $T$ while fixing $k$ reduces the effect of penalization on parameter estimation, and hence on forecasting performance. This in fact can be seen from the tables of the results. Tables 5, 6, 7, 8, 9, and 10 show the performance of the methods when $\eta = 1, 2, 3, 4, 5, 6$, respectively. In this scenario, DHGLASSO for the choice of $\eta = 5$ outperforms the other methods in terms of MPSE/MRPE. Again, DHGLASSO is the most consistent penalty function with respect to the increase in $\eta$.

Table 5: MSPE for October 6$^{th}$ and 7$^{th}$ data combined with $\eta = 1$

| Model | MSPE | MRPE | AIC | BIC |
|---|---|---|---|---|
| VAR | 0.7103 | 14.544 | 204.3445 | 230.15 |
| STAR | 0.253 | 3.9068 | 182.6419 | 183.3035 |

| Model | MSPE | MRPE | AIC | BIC |
|---|---|---|---|---|
| LASSO | 0.2527 | 3.8983 | 182.5923 | 183.254 |
| HGLASSO | 0.2527 | 3.8983 | 182.5923 | 183.254 |
| DHGLASSO | 0.2527 | 3.8983 | 182.5923 | 183.254 |

Table 6: MSPE for October $6^{th}$ and $7^{th}$ data combined with $\eta = 2$

| Model | MSPE | MRPE | AIC | BIC |
|---|---|---|---|---|
| STAR | 0.2273 | 4.0633 | 178.3005 | 179.6239 |
| LASSO | 0.2273 | 4.0633 | 178.3005 | 179.6239 |
| HGLASSO | 0.2273 | 4.0633 | 178.3005 | 179.6239 |
| DHGLASSO | 0.2273 | 4.0633 | 178.3003 | 179.6237 |

Table 7: MSPE for October $6^{th}$ and $7^{th}$ data combined with $\eta = 3$

| Model | MSPE | MRPE | AIC | BIC |
|---|---|---|---|---|
| STAR | 0.2249 | 4.1741 | 177.9721 | 179.9571 |
| LASSO | 0.2248 | 4.1703 | 177.9496 | 179.9347 |
| HGLASSO | 0.2249 | 4.1742 | 177.957 | 179.9421 |
| DHGLASSO | 0.2238 | 4.1062 | 177.6838 | 179.6519 |

Table 8: MSPE for October $6^{th}$ and $7^{th}$ data combined with $\eta = 4$

| Model | MSPE | MRPE | AIC | BIC |
|---|---|---|---|---|
| STAR | 0.2247 | 4.4178 | 178.271 | 180.9008 |
| LASSO | 0.2244 | 4.3892 | 178.0977 | 180.6765 |
| HGLASSO | 0.2247 | 4.419 | 178.2357 | 180.8654 |
| DHGLASSO | 0.2224 | 4.162 | 177.6367 | 180.2156 |

Table 9: MSPE for October $6^{th}$ and $7^{th}$ data combined with $\eta = 5$

| Model | MSPE | MRPE | AIC | BIC |
|---|---|---|---|---|
| STAR | 0.2279 | 3.5851 | 178.7162 | 182.0077 |
| LASSO | 0.2257 | 3.5265 | 178.0827 | 181.1196 |
| HGLASSO | 0.2277 | 3.5857 | 178.6503 | 181.9418 |
| DHGLASSO | 0.2212 | 3.835 | 177.8113 | 180.95 |

Table 10: MSPE for October $6^{th}$ and $7^{th}$ data combined with $\eta = 6$

| Model | MSPE | MRPE | AIC | BIC |
|---|---|---|---|---|
| STAR | 0.2405 | 3.5611 | 178.4606 | 182.4137 |
| LASSO | 0.238 | 3.4261 | 177.7624 | 181.2913 |
| HGLASSO | 0.238 | 3.5376 | 178.2116 | 182.1647 |
| DHGLASSO | 0.2291 | 3.9304 | 178.8057 | 182.4703 |

Another benefit of using STAR-based models in that one can infer the neighborhood influence of other zip-codes demands on a target zip-code. Figures (5), (6), and (7) show the inferred neighborhood correlation among the $\eta = 5$ different neighborhood order for lower, midtown, and upper Manhattan, respectively. The colors on these plots are basically $|\Phi_i|$ for different components of $i$ based on the DHGLASSO method. It's clear from the plots from all lower, midtown, and upper Manhattan, that the correlation/influence between neighboring zip-codes are decreasing as they get farther away from each other. This correlation structure seen in figures (5), (6), and (7) are reasonable and well-aligned with the assumption of using spatio-temporal model, the STARMA model, for predicting taxi demand in Manhattan, New York. In other words, for predicting the taxi demand of the next 15 min for a zip-code in lower Manhattan, the knowledge of the short-term demand history from neighboring zip-codes in lower Manhattan will be more informative as compared to knowing about the short-term demand history of zip-codes in the upper Manhattan. Within STARMA structure, the proposed DHGLASSO model is able to capture this decreasing trend accurately, by reaching the least prediction error among all other methods.

Another notable feature that can be highlighted using the proposed generalized STAR model using DHGLASSO is the variation in the spatial differences in the dependence of demand of neighboring zip-codes. From Figure 5 it can be seen that the value of the coefficients of second and third level of neighboring zip codes is not the same among the zip codes even in lower Manhattan. More specifically, for zip code 10280, the coefficient for the second level neighbors' demand is less than that for the third level neighbor. However, for zip code 10002, the coefficients for first, second and third level neighbors' zip codes demands decrease with level of neighborhood. This non-linear trend of the coefficients for neighboring zip codes could be due to the smaller area of zip codes – particularly for zip codes 10004 and 10280.

*Figure 5: Neighborhood level estimated coefficients for lower Manhattan (zip code: 10004, 10002, 10280)*

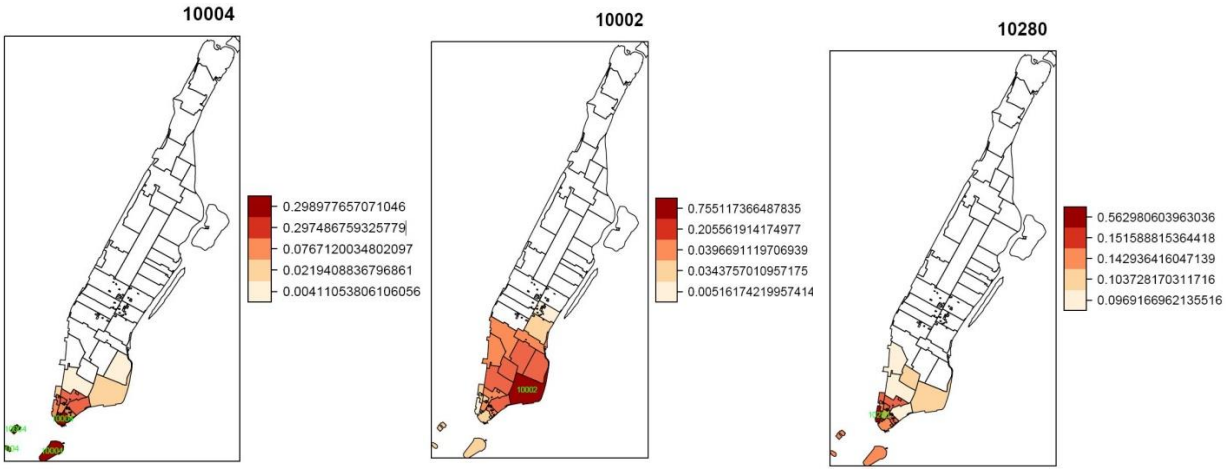

*Figure 6: Neighborhood level estimated coefficients for midtown Manhattan (zip code: 10019, 10022, 10128)*

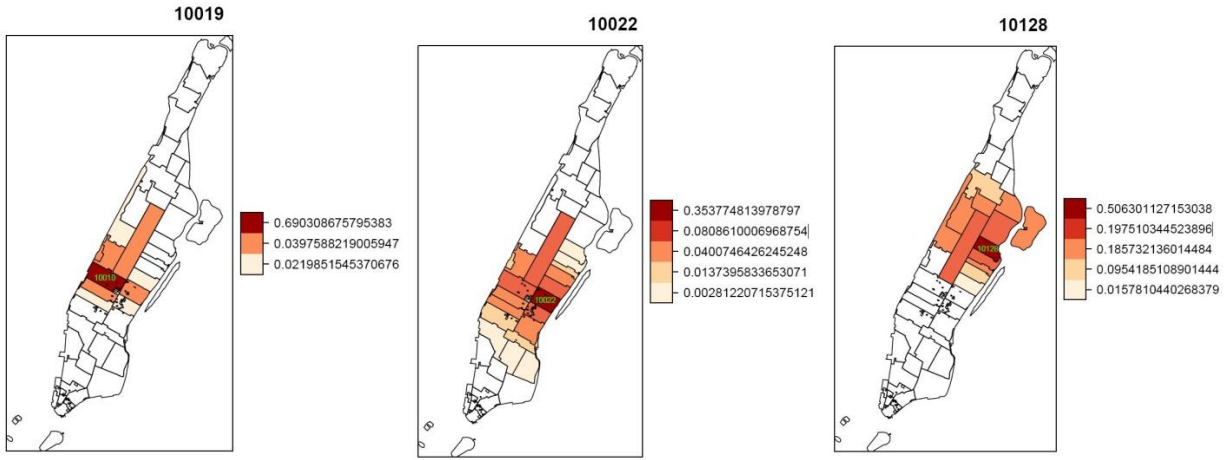

*Figure 7: Neighborhood level estimated coefficients for upper Manhattan (zip code: 10021, 10028, 10027)*

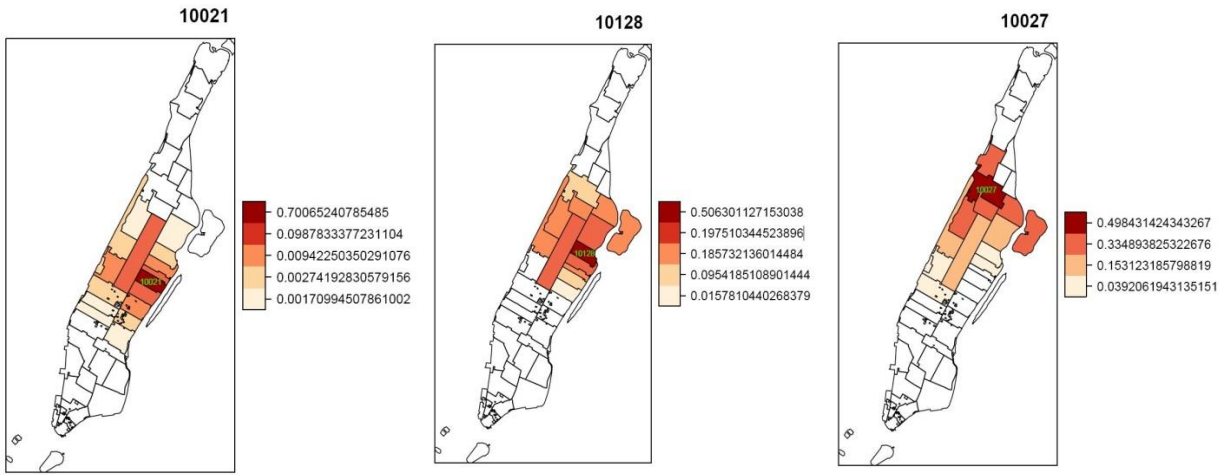

## 5. Conclusion:

Predicting yellow taxi demand in large, populous, and dense area of cities like New York is hard to achieve, since there are numerous parameters affecting its demands. Moreover, in such dense areas, the demand for taxis in different parts of the city are highly variable depending on the time of the day. In this study, taxi demand data obtained from the GPS-enabled historical demand for individual taxis (obtained from NYC TLC) is aggregated spatially by zip-code temporally for every 15-min time interval. A multivariate spatio-temporal method called STARMA is proposed. STARMA reduces the number of parameters dramatically - compared to typical multivariate time series model such as VAR by means of neighborhood structure between the regions. This structure is also useful in capturing the spatial dependence of the demand between the regions and further makes the results more interpretable. Also, a new method for penalizing prediction parameters called double hierarchical group LASS (DHGLASSO) is presented. DHGLASSO penalizes to a larger extent, the parameters that are farther away not only temporally but also spatially – thus establishing a 'double' hierarchy. ()

The proposed model and several other comparable time series models and penalty function are applied over yellow taxi demand of Manhattan for a typical day of the week. The result has revealed that the proposed model could capture the structure of the data well by reaching less prediction error as compared to other time series models such as VAR, STAR with and without LASSO, etc. Using data from both a single day and two consecutive days, proposed generalized STAR model with DHGLASSO performed the best in terms of predictive performance. For the model using data from two consecutive days, a maximum level of neighbors five performed the best. Additionally, DHGLASSO is shown to be most consistent and stable in dealing with increasing parameter dimension.

The proposed generalized STAR model and penalty function is able to capture the spatial variation in the demand for taxis among zip code very well. The effect of neighborhood structure changes depending on the location of interest. The influence of neighborhood taxi demand levels can be easily interpreted – especially by agencies that manage taxi operations and other TNCs. The neighborhood taxi demand dependence can easily be used by taxi companies and TNCs to direct taxi drivers to remain in a certain area depending on the time of day and location. This helps reducing the length of empty taxi trips that seek new rides – thus reducing the emissions, improving air quality and fuel costs for the operators. The computational efficiency due to the DHGLASSO penalization structure helps estimating the model in real-time. Thus, the parameters can be estimated in real-time by agencies such as TLC or TNCs such as Uber or Lyft, which receive the taxi demand data in real-time.

As a part of ongoing and future work, the modeling framework in being extended using other forms of disaggregating. Also, utilizing additional travel demand-related information such as subway and bus ridership, bike demand, weather, etc., will be considered as adding exogenous variables to the time series regimes.

**References:**

1. Baklanov, A.; Hänninen, O.; Slørdal, L.H.; Kukkonen, J.; Bjergene, N.; Fay, B.; Finardi, S.; Hoe, S.C.; Jantunen, M.; Karppinen, A.; et al. (2007). Integrated systems for forecasting urban meteorology, air pollution and population exposure. Atmos. Chem. Phys. 2007, 7, 855–874.

2. Beck, A. & Teboulle, M. (2009). A fast iterative shrinkage-thresholding algorithm for linear inverse problems. SIAM Journal on Imaging Sciences 2(1), 183202.

3. Chang H, Tai Y Hsu JY. (2010). Context-aware taxi demand hotspots prediction. International Journal of Business Intelligence and Data Mining. 5(1): 3–18.

4. Cheng T., Wang J., Harworth J. Heydecker B.G., and Chow A.H.F. (2011). Modeling Dynamic Space-Time Autocorrelation of Urban Tranport Network. GeoComputation, Session 5A: Network Complexity.

5. Correa D., Xie K., Ozbay K. (2017). Exploring the Taxi and Uber Demands in New York City: An Empirical Analysis and Spatial Modeling. Transportation Research Board's 96th, Annual Meeting, Washington, D.C.

6. Duan P., Mao G., Zhang C., and Wang S., (2016). STARIMA-based Traffic Prediction with Time-varying Lags. IEEE 19th International Conference on Intelligent Transportation System (ITSC), Rio, Brazil.

7. Giacinto Di V. (1994). Su una generalizzazione dei modelIi spazio-temporali autoregressivi media mobile (STARMAG). Atti della XXXVII Riunione Scienti_ca SIS, Sanremo, aprile 1994, vol. H.

8. Giacinto Di V. (2006). A generalized space-time ARMA model with an application to regional unemployment analysis in Italy. International Regional Science Review, 29(2), pp.159-198.

9. Giacomini, R. and C. W. J. Granger (2004). Aggregation of space-time processes, Journal of Econometrics 118, 7–26.

10. Jenatton, R., Mairal, J., Obozinski, G. & Bach, F. (2011). Proximal methods for hierarchical sparse coding. The Journal of Machine Learning Research 12, 22972334.

11. Kamarianakis Y. and Prastacos P. (2003). Forecasting traffic flow conditions in an urban network: comparison of multivariate and univariate approaches. Transportation Research Record: Journal of the Transportation Research Board, (1857):74–84.

12. Kamarianakis Y., Prastacos P., and Kotzinos D. (2004). Bivariate traffic relations: A space-time modeling approach. AGILE proceedings, pages 465–474.

13. Kyriakidis, P. C. and A. G. Journel (1999). Geostatistical space-time models: a review, Mathematical Geology 31, 651–683.

14. Lutkepohl, H. (2007). *New introduction to multiple time series analysis*. Springer.

15. Min X., Hu J., Chen Q., Zhang T., and Zhang Y. (2009). Short-Term Traffic Flow Forecasting of Urban Network Based on Dynamic STARIMA Model. Proceedings of the 12th International IEEE Conference on Intelligent Transportation Systems, St. Louis, MO, USA, October 3-7.

16. Min X., Hu, J., and Zhang Z. (2010). Urban Traffic Network Modeling and Short-term Traffic Flow Forecasting Based on GSTARIMA Model. 13th International IEEE Annual Conference on Intelligent Transportation Systems Madeira Island, Portugal, September 19-22.

17. Moghimi, B., Safikhani, A., Kamga, C., & Hao, W. (2017). Cycle Length Prediction in Actuated Traffic Signal Control Using ARIMA Model. ASCE Journal of Computing in Civil Engineering, (In Press).

18. Moreira-Matias L., Gama J., Ferreira M., Mendes-Moreira J., and Damas, L. (2013). Predicting Taxi-Passenger Demand using Streaming Data. IEEE Transactions on Intelligent Transportation Systems, Volume 14, Issue: 3, DOI: 10.1109/TITS.2013.2262376

19. New York City Taxi & Limousine Commission. (2016). TLC Factbook, http://www.nyc.gov/html/tlc/downloads/pdf/2016_tlc_factbook.pdf

20. Nicholson, W.B., Bien, J. and Matteson, D.S., (2014). Hierarchical vector autoregression. arXiv preprint arXiv:1412.5250.

21. Nicholson, W.B., Matteson, D.S., and Bien, J., (2017). VARX-L: Structured Regularization for Large Vector Autoregressions with Exogenous Variables. arXiv preprint arXiv:1508.07497.

22. NYC Taxi & Limousine Commission, http://www.nyc.gov/html/tlc/html/about/factbook.shtml

23. Okutani I. and Stephanedes Y.J. (1984). Dynamic prediction of traffic volume through kalman filtering theory. Transportation Research Part B: Methodological, 18(1):1–11.

24. Pfeifer, P. E., & Deutrch, S. J. (1980). A three-stage iterative procedure for space-time modeling phillip. Technometrics, 22(1), 35-47.

25. Pfeifer, P. E., & Deutsch, S. J. (1981). Variance of the sample space-time autocorrelation function. Journal of the Royal Statistical Society. Series B (Methodological), 28-33.

26. Qian X., Ukkusuri S.V., and Yang C., Yan F. (2017). A Model for Short-Term Taxi Demand Forecasting Accounting for Spatio-Temporal Correlations. Transportation Research Board Annual 2017, Washington D.C.

27. Sartoris, A. (2005). A STARMA model for homicides in the city of Sao Paulo, Proceedings of the Spatial Economics Workshop, Kiel Institute for World Economics, 8–9 April, 2005, Kiel, Germany.

28. Sayarshad HR. & Chow J.Y.J. (2016). Survey and empirical evaluation of nonhomogeneous arrival process models with taxi data, Journal of Advanced Transportation, vol. 50, pp. 1275–1294. DOI: 10.1002/atr.1401

29. Schaller B. (1999). Elasticities for taxicab fares and service availability. Transportation, 26:283–297, DOI: 10.1023/A:1005185421575

30. Schaller B. A. (2005). regression model of the number of taxicabs in U.S. cities. Journal of Public Transportation, 8:63–78, DOI: http://dx.doi.org/10.5038/2375-0901.8.5.4

31. Song, S. & Bickel, P. (2011). Large vector auto regressions. journal: arXiv preprint arXiv:1106.3915.

32. Stathopoulos A., and Karlaftis M.G. (2003). A multivariate state space approach for urban traffic flow modeling and prediction. Transportation Research Part C: Emerging Technologies, 11(2):121–135.

33. Terzi, S. (1995). Maximum likelihood estimation of a generalized STAR(p, lp) model. Journal of the Italian Statistical Society, vol. 4, n. 3, 1995.

34. Tibshirani, R., (1996) Regression shrinkage and selection via the lasso. Journal of the Royal Statistical Society. Series B (Methodological), pp.267-288.

35. Yang C., Gonzales E. (2017). Modeling Taxi Demand and Supply in New York City Using Large-Scale Taxi GPS Data. Seeing Cities Through Big Data - Research, Methods and Applications in Urban Informatics, pp 405-425, DOI10.1007/978-3-319-40902-3_22.

36. Yuan J, Zheng Y, Zhang L, Xie X, Sun G. (2011). Where to find my next passenger? Proceeding UbiComp, New York, NY, 2011.